\input harvmac.tex
\input epsf.tex

\noblackbox
\def\sq2{\sqrt{2}}
\def\ra{\rightarrow}

\def\p{\partial}

\def\RN{Reissner-Nordstr\"om}

\def\s42{ 2^{-{1\over 4} } }

\def\ra{\rightarrow}

\def\cg{\cos (2\pi V)}
\def\sg{\sin (2\pi V)}
\def\cg2{\cos (\pi V)}
\def\sg2{\sin (\pi V)}
\def\cb2{\cos (\delta/2)}
\def\sb2{\sin (\delta/2)}

\font\cmss=cmss10 \font\cmsss=cmss10 at 7pt
\def\IZ{\relax\ifmmode\mathchoice
   {\hbox{\cmss Z\kern-.4em Z}}{\hbox{\cmss Z\kern-.4em Z}}
   {\lower.9pt\hbox{\cmsss Z\kern-.4em Z}}
   {\lower1.2pt\hbox{\cmsss Z\kern-.4em Z}}\else{\cmss Z\kern-.4emZ}\fi}

\lref\hs{G. Horowitz and A. Strominger, Nucl.Phys {\bf B 360} (1991) 197.}

\lref\sv{A. Strominger and C. Vafa, hep-th/9601029.}

\lref\cmp{C.C. Callan, J.M. Maldacena and A.W. Peet, hep-th/9510134,
to appear in Nucl. Phys. B.}

\lref\tseytlin{ A.A. Tseytlin, hep-th/9601177; M. Cvetic and A.A. Tseytlin,
hep-th/9512031. }

\lref\tseylinother{ M. Cvetic and A. Tseytlin, hep-th/9510097; 
G. Horowitz and A. Tseytlin, Phys.Rev.{\bf D51} (1995) 2896, hep-th/9409021.}

\lref\ckmy{C.G. Callan, I.K. Klebanov, J.M. Maldacena and A. Yegulalp,
Nucl.Phys.{\bf B443} (1995) 444, hep-th/9503014.}

\lref\sentduality{ A. Sen, hep-th/9512203.}

\lref\dh{A. Dabholkar and J. Harvey, Phys. Rev. Lett. {\bf 63} (1989) 478;
A. Dabholkar, G. Gibbons, J. Harvey and F. Ruiz-Ruiz, Nucl. Phys. {\bf B340}
(1990) 33; A. Dabholkar, 
J. Gauntlett, J. Harvey and D. Waldram, hep-th/9511053.}

\lref\callanfive{ C. Callan, J. Harvey and A. Strominger, 
Nucl. Phys. {\bf B359 }
(1991) 611. }

\lref\larsen{ F. Larsen and F. Wilczek, hep-th/9511064.}

\lref\myresperry{ F. Tangherlini, Nuovo Cimento {\bf 77} (1963) 636;
R. Myers and M. Perry, Ann. Phys. {\bf 172} (1986) 304.}

\lref\polchinski{ J. Dai, R. Leigh and J. Polchinski, Mod. Phys. Lett.
{\bf A 4} (1989) 2073; P. Horava, Phys. Lett. {\bf B231}
(1989) 251; J. Polchinski, hep-th/9510017.}

\lref\cghs{C. Callan, S. Giddings, J. Harvey and
A. Strominger, Phys. Rev. {\bf D45} (1992) 403.}

\lref\limitations{
J. Preskill, P. Schwarz, A. Shapere, S. Trivedi and F. Wilczek,
Mod. Phys. Lett.{\bf A 6} (1991) 2353; P. Kraus and F. Wilczek, 
Nucl. Phys. {\bf B433} (1995) 403. }

\lref\andyunitarity{ A. Strominger, hep-th/9410187; J. Polchinski and
A. Strominger, Phys.Rev. {\bf D50} (1994) 7403,  hep-th/9407008 
}

\lref\schwartzmultstrings{
 J. Schwarz, Phys.Lett. {\bf B360} (1995) 13; 
ERRATUM-ibid.{\bf B364} (1995) 252.}

\lref\hull{
C. Hull and P. Townsend, Nucl. Phys. {\bf B438} (1995) 109, 
hep-th/9410167.}

\lref\horavad{ P. Horava, Phys. Lett. {\bf B 289} (1992) 293.}

\lref\schwarzhair{ J. Schwarz, Phys. Lett. {\bf 272B} (1991) 239.}

\lref\susskindu{ L. Susskind and J. Uglum, Phys. Rev. {\bf D 50}
 (1994) 2700.}

\lref\wittenbound{E. Witten, hep-th/9510135.}

\lref\vafacount{ C. Vafa, hep-th/9511088; C. Vafa, hep-th/9512078}

\lref\wittensmall{ E. Witten, hep-th/9511030.}

\lref\complementarity{ G. 't Hooft, Nucl.Phys {\bf B 335} (1990) 138;
L. Susskind, L. Thorlacius and J. Uglum, {\bf D 48} (1993) 3743;
Y. Kiem, H. Verlinde and E. Verlinde Phys.Rev.{\bf D52}, 
 (1995) 7053 hep-th/9502074; D. Lowe, J. Polchinski, L. Susskind, 
L. Thorlacius and J. Uglum, Phys.Rev. {\bf D52} (1995) 6997,
 hep-th/9506138; V.  Balasubramanian and H. Verlinde,
hep-th/9512148. }

\lref\hawkingt{ S.W. Hawking, Comm. Math. Phys. {\bf 43} (1975) 199.}

\lref\membrane{ K. Thorne, R. Price and D. MacDonald,`` {\it
Black Holes: The Membrane Paradigm }'', Yale University Press, 1986;
L. Susskind, L. Thorlacius, J. Uglum 
Phys. Rev. {\bf D48} (1993) 3743, hep-th/9306069 
 }

\lref\hawkingentropy{ J. Beckenstein, Phys. Rev. {\bf D7} (1973) 2333;
Phys. Rev. {\bf D 9} (1974) 3292; S. W. Hawking, Phys. Rev. {\bf D13 }
(1976) 191.}

\lref\hawkinguni{ S. Hawking, Phys. Rev. {\bf D 14} (1976) 2460; 
For a recent review see S. Giddings, hep-th/9508151 and references therein.}

\lref\hologram{ L. Susskind, J. Math. Phys. {\bf 36 } (1995) 6377;
L. Susskind, J. Math Phys. {\bf 36} (1995) 6377. }

\lref\lenyspeculations{ L. Susskind, RU-93-44, hep-th/9309145.}

\lref\scattering{ I. Klebanov and L. Thorlacius, hep-th/9510200, to appear
in Phys. Lett. B; S. Gubser, A. Hashimoto, I. Klebanov and J. Maldacena,
 hep-th/9601057.}

\lref\vafaintersecting{ M. Bershadsky, C. Vafa and V. Sadov, 
hep-th/9510225; A. Sen, hep-th/9511026.}

\lref\dasmathur{  S.R. Das and  S.D. Mathur, 
hep-th/9601152.}

\lref\wittenpol{ J. Polchinski and E. Witten, hep-th/9510169.}

\lref\dimensionalreduction{
J. Maharana and J. Schwarz, Nucl.Phys. {\bf B390} (1993) 3; 
A. Sen, Nucl. Phys. {\bf D404} (1993) 109.}
 
\lref\cvyo{ M. Cvetic and D. Youm,
 hep-th/9510098.} 

\lref\topological{ S. Ferrara, R. Kallosh and A. Strominger, Phys.
Rev. {\bf D 52} (1995) 5412, hep-th/9508072; M. Cvetic and D. Youm,
hep-th/9507090; G. Gibbons and P. Townsend, Phys. Rev. Lett. {\bf 71}
(1993) 3754.}

\lref\polchinskinotes{ J. Polchinski, S. Chaudhuri and C. Johnson,
hep-th/9602052.}

\lref\hstro{ G. Horowitz and A. Strominger, hep-th/9602051}

\Title{\vbox{\baselineskip12pt
\hbox{PUPT-1591}\hbox{hep-th/9602043}}}
{\vbox{\centerline{\bf  D-Brane Approach to Black Hole Quantum Mechanics}}}
\centerline{Curtis G. Callan, Jr. and  Juan M. Maldacena \footnote{$^\dagger$}
{e-mail addresses: callan, malda@puhep1.princeton.edu }
}
\centerline{\it Department of Physics, Princeton University}
\centerline{\it Princeton, NJ 08544, USA}
\vskip.3in
\centerline{\bf Abstract}

Strominger and Vafa have used D-brane technology to identify and precisely count
the degenerate quantum states responsible for the entropy of certain 
extremal, BPS-saturated black holes. Here we give a Type-II  D-brane
description of a class of extremal {\it and} non-extremal five-dimensional 
Reissner-Nordstr\"om  solutions and identify  a corresponding set of
degenerate D-brane configurations. We 
use this information to do a string theory calculation of the 
entropy, radiation rate and ``Hawking'' temperature. The results agree 
perfectly with standard Hawking results for the corresponding 
nearly extremal Reissner-Nordstr\"om black holes.  Although these 
calculations suffer from open-string strong coupling
problems, we give some reasons to believe that they are nonetheless
qualitatively reliable. In this optimistic scenario there would be
no ``information loss'' in black hole quantum evolution.

\smallskip

\Date{2/96}
\eject

\newsec{Introduction and Motivation}

Black hole thermodynamics \refs{ \hawkingt,\hawkingentropy} 
implies that black holes have an entropy 
proportional to the area of the event horizon: $S = A /4 G_N$,
where $G_N$ is the Newton constant and $\hbar =1$. Associated with this 
entropy are apparent problems of unitarity and information loss in the
quantum dynamics of the black hole \hawkinguni. It has long been felt that the
resolution of the paradoxes must lie in a microscopic quantum description 
of the states responsible for the entropy and that string theory, 
the only consistent quantum gravity, is the right framework in which to 
look for such a description \refs{\schwarzhair, \susskindu,
\lenyspeculations  }. Progress along this line was, however, held up 
for some time by the absence of sufficiently explicit string theory 
descriptions of extended spacetime objects like macroscopic black holes.
This obstacle has recently been rendered less daunting
by the advent of D-branes \polchinski, which 
give a simple description of stringy solitons in terms of open strings
with unusual boundary conditions. 

Most D-branes are not black holes in any useful sense, but a few examples 
which are have now been found and the problem of counting states has begun 
to receive attention. In particular, Strominger and Vafa \sv\ have found a
D-brane description of an extremal, BPS-saturated, \RN\  black hole in
type-II string theory compactified to five dimensions on $S_1\times K3$
and shown for the first time that state counting reproduces the
Bekenstein-Hawking entropy. Surprisingly detailed features of the 
emerging picture of black hole entropy had actually been anticipated  
in work based on rather general string theory considerations \larsen, and 
it now seems that some of the same features can be extracted from a more 
``conventional'' conformal field theory approach \tseytlin. 

Nonetheless, the D-brane approach represents a major advance 
in our quantitative understanding of these matters and, in this paper, 
we will use it to study  the physics of a class of nearly extremal 
five-dimensional \RN\  black holes. The extremal 
case is very similar to that  studied in \sv\ , but the geometric picture
used here, based on toroidal compactification, seems 
particularly well suited to discussing the non-extremal case.
We work in the type IIB string theory on $ M^5 \times T^5$ and we 
construct a D-brane configuration, preserving 1/8 of
the supersymmetries,  such that the corresponding supergravity
solutions describe black holes with finite horizon area in five dimensions. 
This configuration consists of a large number of 5D-branes wrapped
on $T^5$, plus 1D-branes wrapped along one of the compact directions,
plus an ensemble of open strings with one end attached to a
 1D-brane and the other to a 5D-brane. All the open strings move
in the same direction so that the solution carries a large total 
momentum in this internal dimension. 

Following \sv, we derive the area law  for the  entropy  of the
extremal black hole by state counting. Then we  consider ``nearly'' 
extremal configurations whose deviations from extremality are small 
in the macroscopic sense ( $\delta M /M \ll 1$) but large in the 
microscopic sense ( $\delta M \gg $ masses of low energy excitations). 
A D-brane calculation of entropy  gives exactly the same results as 
the canonical calculation based on the associated \RN\  metric 
\refs{ \hawkingentropy , \myresperry }~ (at least in the leading
nearly extremal limit). Hawking radiation is seen as a simple decay
process with a rate that could be  calculated using  perturbative
string techniques. This rate has thermal properties with 
exactly the Hawking temperature \hawkingt . The overall coefficient
is proportional to the area.

A skeptic could object to our naive perturbative calculation on the 
grounds that the open strings attached to a macroscopic horizon are 
inevitably strongly coupled. We will present qualitative arguments 
why the classical black hole description and the perturbative D-brane 
picture could be valid at the same time, thus rationalizing 
our treatment of non-extremal black hole physics. In this optimistic 
scenario, quantum evolution of the D-brane system would be  
manifestly unitary and there would  be no black hole information loss.

An essential issue, which is brought to the fore by our treatment,
is the manner in which the states responsible for the black hole
entropy encode, as they must, the state of particles which have fallen
through the macroscopic horizon. If this can be done, it may give 
give concrete shape to the imaginative ``holographic'' ideas about
horizon dynamics that have been proposed by Susskind
\refs{\susskindu ,\lenyspeculations}. 


\newsec{Classical Black Hole Solutions in 5 Dimensions.}

The five dimensional \RN\ black hole is a solution of
the five dimensional Einstein plus Maxwell action. 
The metric reads \myresperry 
\eqn\reno{
ds^2 = - \lambda dt^2 + \lambda^{-1} dr^2 + r^2 d\Omega_3^2 }
$$ \lambda = \left(  1-  {r_+^2 \over r^2}  \right) 
\left(  1-  {r_-^2 \over r^2}  \right)
$$
There is a horizon at 
 $ r = r_+ $,
mass and charge are given by\footnote{$^\dagger$}{Here, and
in the rest of the paper, $G_N$ denotes the five dimensional
Newton constant.}
\eqn\mass{
 M = { 3 \pi \over 8 G_N } ( r_+^2 + r_-^2) ~~~~~~~~~~
 Q = r_+ r_-
}
and the entropy is proportional to the area of the
horizon, $ S = {A /  4 G_N} =  \pi^2 r_+^3/ 2 G_N  $.
The extremal solution is obtained by taking $r_+ = r_-\equiv r_e $ and has
zero temperature but non-zero entropy. If we take a nearly 
extremal solution ($ r_+ \sim r_- $), keeping the charge
fixed, we find that the entropy behaves as 
\eqn\entropy{ 
{ \Delta S \over S_e } = { 3 \over \sqrt{2}} 
\sqrt{ \Delta M \over M_e }
}
where $M_e$, $S_e$ are the extremal values of 
the mass and entropy respectively. The Hawking temperature for
such a nearly extremal black hole is
\eqn\hawking{ 
T_H =  { 2 \over \pi r_e }  \sqrt{ \Delta M \over 2 M_e } 
} 


Now let us show that there is a configuration of strings
and solitons in the type IIB string theory on $M^5 \times T^5$
 that can be  be identified with 
the \RN\ black hole. 
We denote the $M^5$ coordinates by $(t, x_1, x_2,x_3,x_4)$ and
the torus coordinates by $(x_5,x_6,x_7,x_8,x_9)$.
The  configuration we study is constructed as follows:
 first wrap $Q_5$ 5D-branes on the torus, then 
wrap $Q_1$ 1D-branes along an $S_1$ of the torus  
(to be definite, let us say that it is the   direction $\hat 5$).
This leaves $1/2 \times 1/2 = 1/4$ of the supersymmetries unbroken. 
To see this, notice that 
in  the ten dimensional type IIB theory the supersymmetries
were generated by  two independent chiral spinors $\epsilon_R$ 
and $\epsilon_L$ ( $\Gamma^{11} \epsilon_{R,L} =\epsilon_{R,L}$). 
The presence
of the strings and the fivebranes imposes  additional conditions
on the surviving supersymmetries 
\eqn\susyonefive{
 \epsilon_{R} = \Gamma^0 \Gamma^5 \epsilon_L ~~~~~~~~~
 \epsilon_{R} = \Gamma^0 \Gamma^5 \Gamma^6\Gamma^7\Gamma^8\Gamma^9 
 \epsilon_L
}
If, in addition, we put some left-moving momentum along 
the $\hat 5$  direction
then we  further break the supersymmetry to $1/8$ of the original, 
through  the extra condition
\eqn\extra{ 
\Gamma^0 \Gamma^5 \epsilon_R =  \epsilon_R.
}
Taken together, this gives 
 the following decomposition
for the surviving  spinor
\eqn\spinor{ 
\epsilon_L = \epsilon_R= 
\epsilon_{SO(1,1)}^+ \epsilon_{SO(4)}^+ \epsilon_{SO(4)}^+
}
The positive chirality $SO(4)$ spinor is pseudoreal and has two independent
components so that 4 out of the original 32 supersymmetries are preserved
by this configuration.
Now we want to find a classical supergravity 
solution associated to
this 1-brane plus 5-brane plus momentum configuration. 
Fortunately, Tseytlin \tseytlin\ has found a closely related
classical solution in type II string theory on $M^5 \times T^5$. 
His configuration consists of solitonic five branes and
fundamental strings, and carries momentum along the direction $\hat 5$. 
In order to find a D-brane description we use the 
strong-weak coupling duality 
of this theory {\refs{\schwartzmultstrings ,\hull }}.
 Such a duality transformation turns fundamental strings 
into 1D-branes and solitonic 5-branes into 5D-branes and turns Tseytlin's
solution into just what we are looking for.

The dilaton field  and the 
 ten dimensional string metric emerging from this procedure are
$$ e^{- 2 \phi_{10} } =  { f_5 ~ f_1^{-1} } $$
\eqn\metric{ \eqalign{ d s^2_{str} = &  f^{-{1\over 2}}_1 
f^{-{1\over 2}}_5
 \left( -dt^2 + dx_5^2 + 
K (dt-dx_5)^2 \right) + \cr
 &f^{{1\over 2}}_1 f^{{1\over 2}}_5 ( dx_1^2 +  \cdots + dx^2_4) +
  f^{{1\over 2}}_1 f^{-{1\over 2}}_5 ( dx_6^2  + \cdots + dx^2_9 ).} ~ }
The solution contains three harmonic functions of the transverse coordinates 
($x^2 = x_1^2 +  \cdots + x^2_4 $) associated with the three charges 
needed to define the solution:
\eqn\harmonic{ f_1 = 1 + { c_1 Q_1 \over x^2 }~~~~~~~~~~~~~~~~~
f_5 = 1 + { c_5 Q_5 \over x^2 } ~~~~~~~~~~~~~~~~~~
K =  { c_p N  \over x^2 }
}
Some components of the Ramond-Ramond antisymmetric
tensor field, $B^{RR}_{\mu\nu}$,
are also excited and they     behave as gauge fields when
we dimensionally reduce to five dimensions.
The three independent charges arise as follows:
$Q_1$ is a RR electric charge, coming from $B^{RR}_{05}$ and
counts the 1D-branes.
 $Q_5$ is a magnetic charge for the
three form field strength $H^{RR}_3 = d B^{RR}_2 $, which is
dual in five dimensions to a gauge field, $ H^{RR}_3=* F_2 $.
 $Q_5$ is thus an
electric charge for the gauge field $F_2$ and it counts the 
5D-branes.
The third charge, $N$, corresponds to 
 the total momentum carried by the open strings traveling
on the branes in the direction $\hat 5$, 
and  it is associated
to the five dimensional 
Kaluza-Klein
gauge field coming from the $G_{50}$ component
of the metric. 
The coefficients in \harmonic ,
\eqn\coeff{
c_1 = { 4  G_N R  \over g \pi \alpha'}~~~~~~~~~~~~~~~~~
c_5 =  \alpha' g ~~~~~~~~~~~~~~~~~
c_p = {  4  G_N \over  \pi R},
}
are defined so that the charges $Q_1$, $Q_5$ and $N$ are naturally
integer quantized. In these expressions, $g $ is the ten dimensional
string coupling 
constant, defined so that S-duality takes $ g\ra 1/g$ (note that 
$G_N \sim g^2$\footnote{$^\dagger$}{The precise relation
between g and $G_N$ will not be need in what follows.}).
The precise value of these coefficients follows from matching the 
field of a fundamental string to its source following \dh\ and 
from the solitonic fivebrane Dirac quantization for the charge \callanfive .
Note that the total momentum in the direction $\hat 5$ is quantized as
$N/R$, where $R$ is the radius of the circle.

Now we dimensionally reduce \metric ~to five dimensions in order to
read off black hole properties.
The standard method of  \dimensionalreduction ~yields
a five-dimensional Einstein metric, 
$ g^5_E = e^{ - 4 \phi_5/3 } G^5_{string} $,
\eqn\einstein{
ds^2_E = - { 1 \over
\left( f_1 f_5 (1+K) \right)^{2 \over 3} } dt^2 +
\left( f_1 f_5 (1+K) \right)^{1 \over 3}
( dx_1^2 + \cdots + dx_4^2 )
}
which can be interpreted as a five dimensional extremal, charged, 
supersymmetric black hole with nonzero horizon area. Calculating the 
horizon area in this metric \einstein ~we get the entropy
\eqn\entropyext{
S_e  = 
 {A_H \over 4 G_N } =  2 \pi \sqrt{N Q_1 Q_5 }
}
In this form the entropy does not depend on any of the continuous
parameters like the coupling constant or the sizes of the internal
circles, etc. This ``topological'' character of
the entropy was emphasized in  \refs{\topological , \larsen ,\tseytlin}.
 It is also symmetric under interchange of
$N, Q_1, Q_5$. In fact, U duality  \refs{\hull , \sentduality ,
\vafaintersecting } interchanges the three  charges. 
 To show it in a more specific fashion, 
 let us define T$_i$ to be the usual T-duality that inverts the
 compactification radius in the direction $i$ 
 and S the ten dimensional S duality of 
 type IIB theory. Then a transformation that 
sends ($N,Q_1, Q_5$) to ($Q_1, Q_5, N$)
 is U=T$_9$T$_8$T$_7$T$_6$ST$_6$T$_5$. If the ten dimensional 
coupling constant is small and the radii are of order $\alpha'$
 it leaves the
 ten dimensional and the 5 dimensional coupling constants small but
 it modifies the radii of the compact dimensions.
 This symmetry of the string theory will be crucially used below. 
  
The standard five-dimensional extremal \RN\ solution \reno\ is recovered 
when the charges are chosen such that 
\eqn\balance{
c_p N = c_1 Q_1=c_5 Q_5 =r_e^2~.
}
and the coordinate transformation $r^2 = x^2 + r_e^2$ is applied.
The crucial point is that, for this ratio of charges, the dilaton field 
and the internal compactification geometry are independent of position
and the distinction between the ten-dimensional and five-dimensional
geometries evaporates. What is at issue is not so much the charges
as the different types of energy-momentum densities with which they are 
associated. An intuitive picture of what goes on is this: a p-brane produces
a dilaton field of the form $ e^{- 2 \phi_{10} } = f_p^{p-3 \over 2}$,
with $f_p$ a harmonic function \hs. A superposition of branes produces
a product of such functions and one sees how 1-branes can cancel
5-branes in their effect on the dilaton. A similar thing is true for
the compactification volume: For any p-brane, the string metric is such 
that the volume parallel to the brane shrinks and the volume perpendicular 
to it expands as we get closer to the brane. It is easy to see how
superposing 1-branes and 5-branes can stabilize the volume in the directions 
$ \hat 6, \hat 7, \hat 8, \hat 9$, since it is perpendicular to the 1-brane and 
parallel to the 5-brane. The volume in the direction $\hat 5$ would still
seem to shrink, due to the tension of the branes. This is indeed why 
we put momentum along the 1-branes, to balance the tension and produce a stable 
radius in the $\hat 5$ direction. If we balance the charges precisely 
\balance ~(we can always do 
this for large charges) the moduli scalar fields associated 
with the compactified dimensions are not excited at all, which is what we need 
to get the \RN\ black hole. 

\newsec{D-Brane Description of Extremal Black Holes}

Now we want to give a D-brane description of this black 
hole\footnote{$^\dagger$}{For a discussion of a two dimensional
black hole using D-branes see \horavad.}. We
consider type IIB string theory compactified on $M^5\times T^5$. As
we know, type IIB has $2p+1$-brane soliton solutions which have a D-brane
description via open strings with Dirichlet boundary conditions.
In particular, there are both 1-branes and 5-branes and we can use
them to construct a multi-brane state in the same spirit as the
construction of 5D \RN\ black holes in the previous section. To be
precise, we superpose $Q_5$ 5D-branes wrapping around the $T^5$ and $Q_1$ 
1D-branes wrapping around one of the compact directions (call it $\hat 5$).
The supersymmetry analysis is the same as for the classical solution and the
branes with no excitations break 1/4 of the supersymmetries. 
These membranes can have excitations and  we will be interested in
the excitations that  break just one additional supersymmetry by
imposing the extra condition \extra. 
This rules out the massive open strings attached to the D-branes, so
we concentrate then only on the massless excitations of the 
open strings that live on the D-brane. 
Actually \extra  ~implies that the open strings 
have to move in the direction $\hat 5$ and be left moving.
Momentum along the direction  $\hat 5$ is quantized in 
units of $1/R$. If we consider a state containing $n_i $ open strings
with momentum $i$ (all momenta of the same sign!), the total momentum 
in the direction $\hat 5$ is
$$ P = {1 \over R } \sum_i i n_i  = {N \over R }, $$
an expression with the structure of a string level number (this fact was
exploited in a related context in \dasmathur\ to count BPS and non-BPS 
states of oscillating D-strings).

We will first count, via a straightforward (modulo one subtlety) denumeration 
of the open string states supported by this collection of branes,
the degeneracy of the states corresponding to extremal black holes.
Our result for this will, in the end, not be significantly different 
from that of \sv, which is based on a sophisticated analysis of the 
cohomology of instanton moduli spaces \vafacount. We present our own 
method anyway, if only because we find it easier to generalize to the
counting of non-extremal states, the ultimate topic of this paper. 
There are many types of open strings to consider: those that 
go from one 1-brane to another 1-brane, which  we denote  as (1,1) strings, 
as well as the corresponding  (5,5), (1,5) and (5,1) strings (the last two 
being different because the strings are oriented). We want to construct the 
BPS states that have the highest degeneracy and that corresponds to 
taking the strings all of (1,5) or (5,1) type. In fact, it was shown in  
\vafaintersecting\ that, in the low energy effective world volume field 
theory, these states are charged matter fields in the fundamental 
representation of the U($Q_1$)$\times$U($Q_5$) gauge theory generated
by the (1,1) and (5,5) strings (see also \wittenbound). The presence of
many open strings effectively gives an expectation value to these fields, 
which then act as Higgs fields and   break the U($Q$) symmetries. In short, 
if we excite many (1,5) or (5,1) strings the (1,1) and (5,5) strings become 
massive and can be dropped from the state counting. This also indicates that 
the configurations under discussion are really bound states, since the 
transverse motion of the branes relative to each other is brought about by 
(1,1) or (5,5) string excitations. 
\vskip 1cm
\vbox{
{\centerline{\epsfxsize=2in \epsfbox{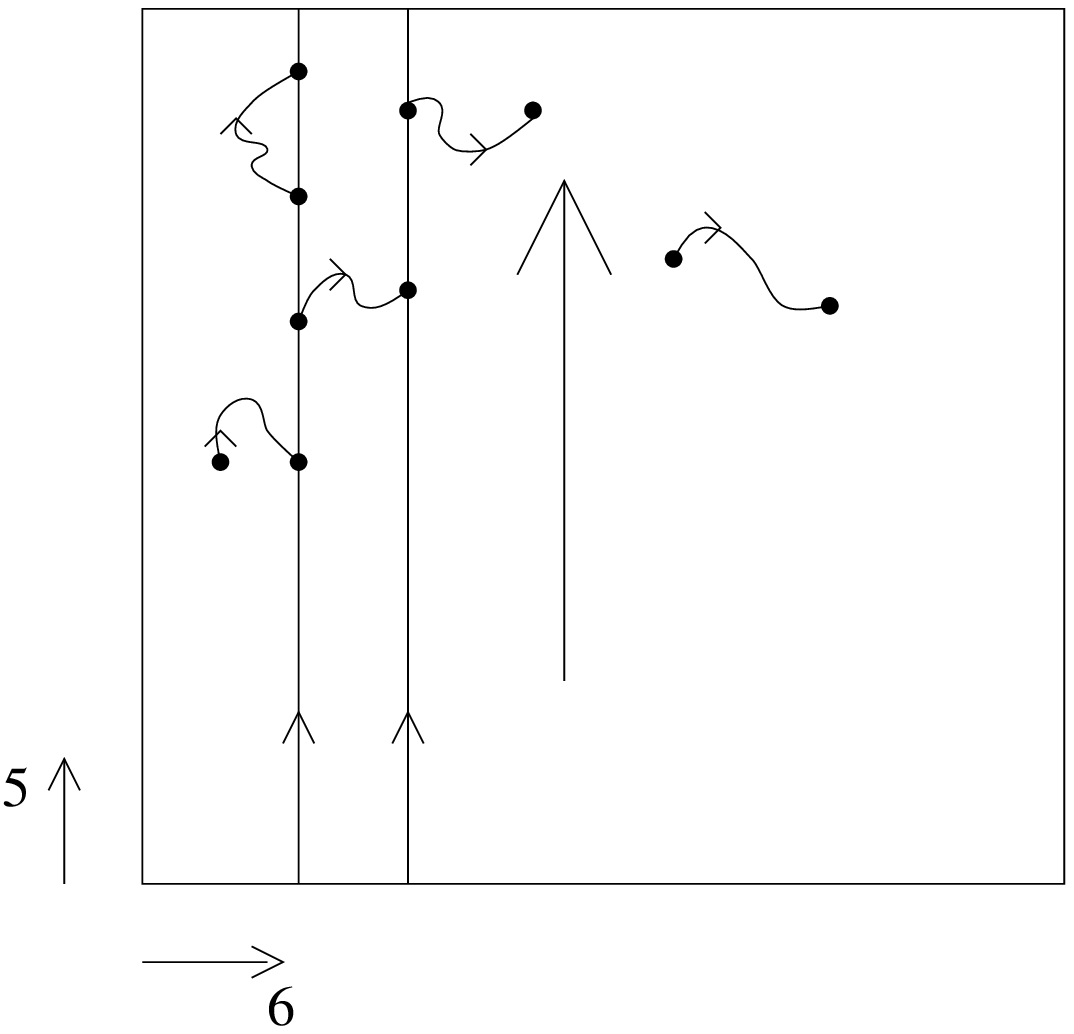}}}
{\centerline{\tenrm FIGURE 1: 
Configuration of intersecting D-branes. We show  two of
the internal
}}
{\centerline{ 
dimensions and several types of open strings. The open 
strings going between
}}
{\centerline{
1 and 5 branes are the most relevant for the black holes
that we analyze.
}}
}
\vskip .5cm

Let us consider then the (1,5) and (5,1) strings.
We have 2 bosons with NN boundary conditions, 4 with ND and
4 with DD. The vacuum energy of the bosons is then
$ E = 4(-1/24 + 1/48)$. Of the fermions in the NS state, the  4 that
are in the ND directions will end up having R-type quantization
conditions. The net fermionic vacuum energy is $ E= 4( 1/24 -  1/48) $ 
and exactly cancels the bosonic one.  
This vacuum is a spinor under SO(4),  is acted on by 
$\Gamma^6,\Gamma^7,\Gamma^8,\Gamma^9
$ and obeys the GSO chirality condition
$ \Gamma^6\Gamma^7\Gamma^8\Gamma^9 \chi = \chi$. What remains is a 
two dimensional representation. There are
two possible orientations and they can be attached to 
any of the different branes of each type. This gives a total
of $ 4 Q_1 Q_5$ different possible states for the string. 
Now consider the Ramond sector. The four internal fermions 
transverse to the string will have NS type boundary conditions. The
vacuum again has zero energy and is an SO(1,5) spinor and thus a  fermion.
Again the GSO condition implies that only the positive
chirality representation of SO(1,5) survives.
It also has to be left moving so that only the $ 2^+_+$ under
SO(1,1)$\times$SO(4) survives. This gives the same number of states as 
for the bosons. In summary, for each momentum we could have 
$4Q_1Q_5$ bosons and  $4Q_1Q_5$ fermions. Roughly speaking, the state 
counting is the same as that of the left moving oscillator modes 
of $4 Q_1 Q_5$  free superconformal fields. The asymptotic formula for 
the degeneracy of level $N$ states in a conformal field theory of 
central charge $c$ is $d(N) \sim e^{ 2 \pi \sqrt{c N/6} }$. Including
both the fermions and the bosons, we have $c= 4 Q_1 Q_5\times 3/2 $ 
leading to an  entropy
\eqn\entropyd{ S_e = 2 \pi\sqrt{ Q_1Q_5 N}~, }
in perfect agreement with \entropyext. This argument is not quite right
since it treats the multiple 1-branes and 5-branes as distinguishable.
Since they are at least separately indistinguishable, one might guess that 
the superconformal fields should be taken to describe the orbifold
$(T^4)^{Q_1 Q_5}/S(Q_1)S(Q_5)$, where $S(Q)$ is the permutation group 
on $Q$ items. A deeper analysis, based on the instanton moduli space
approach \vafacount, strongly suggests the correct answer is
$(T^4)^{Q_1 Q_5}/S(Q_1 Q_5)$ as if, somehow, the 1-branes were effectively 
indistinguishable from the 5-branes as well as themselves. We would like to 
understand this point better\footnote{$^\dagger$}{We are grateful to Cumrum 
Vafa for pointing out an error in our original treatment of this point.} 
but, fortunately for us, the 
various orbifolds have the same central charge and the same large-$N$ behavior
of the degeneracy of states.

It is interesting to note that we could have considered an
analogous black hole in the heterotic string theory on 
$M^5\times T^5$ built out of fundamental strings and solitonic 
five-branes. These are the dyonic black holes considered by 
various authors \refs{\larsen, \tseytlin,\tseylinother, \cvyo}.
In that case, the analogous D-brane description takes place
in the type I theory, which indeed has 1D-branes and 
5D-branes corresponding to the fundamental string and
the solitonic fivebrane \wittenpol. The D-brane counting can also 
be done, and it is interesting to notice that to get the
correct result one must include the 5D-brane SU(2) degrees
of freedom found in \wittensmall. 

Something remarkable has happened here.
We started with some configuration of D-branes sitting at $r=0$, 
a point in 5-dimensional space. To start with, this is a
``point with nothing inside it.'' However, having put all these
open strings on the branes   we find that the configuration 
matches  a solution of  the classical low energy action 
such that  $r=0$ is really  a 3-sphere with non-zero
area! What happened? Well, the ten dimensional classical 
solutions for D-branes show that as we get closer to the D-brane the
transverse space expands and the longitudinal space shrinks. 
This configuration has expanded the transverse space in such
great way that what previously was a point is now a sphere. 
The most exciting aspect of this is that the classical solution 
continues beyond the horizon, into the black hole singularity, whereas,
according to the D-brane picture space simply stops at $r=0$. 
States inside the horizon would have to be described by the massless
modes on the D-brane. The basic horizon degrees of freedom are 
denumerated by three integers: the momentum, the index labeling the 
1-brane and the index labeling the 5-brane. 
When a string ``falls'' into the black hole  and crosses the
horizon, it turns into a pair of  open strings traveling on the 
D-branes (see figure 2). Since an infalling geodesic is labeled by 
three angles on the 3-sphere and the open string states are
labeled by three integers ($i$,$j$,$n$) (with 
$ 1\le i \le Q_1$, $ 1\le j \le Q_5$ and $n$ the momentum), there
is at least a correspondence between numbers of degrees of freedom. 
Of course, the transformation of ``physical'' space coordinates
to the three integers could be very complicated indeed. 
The existence of an upper bound on these integers could be related 
to the fact that there can be at most ``one bit per unit area'' \hologram .
All of this is reminiscent  of the ``holographic'' principle \hologram, 
as well as the membrane paradigm \refs{\lenyspeculations,\membrane},
in that dynamics occurring inside the black hole would be described 
as occurring on the horizon. 

\vskip 1cm
\vbox{
{\centerline{\epsfysize=2in \epsfbox{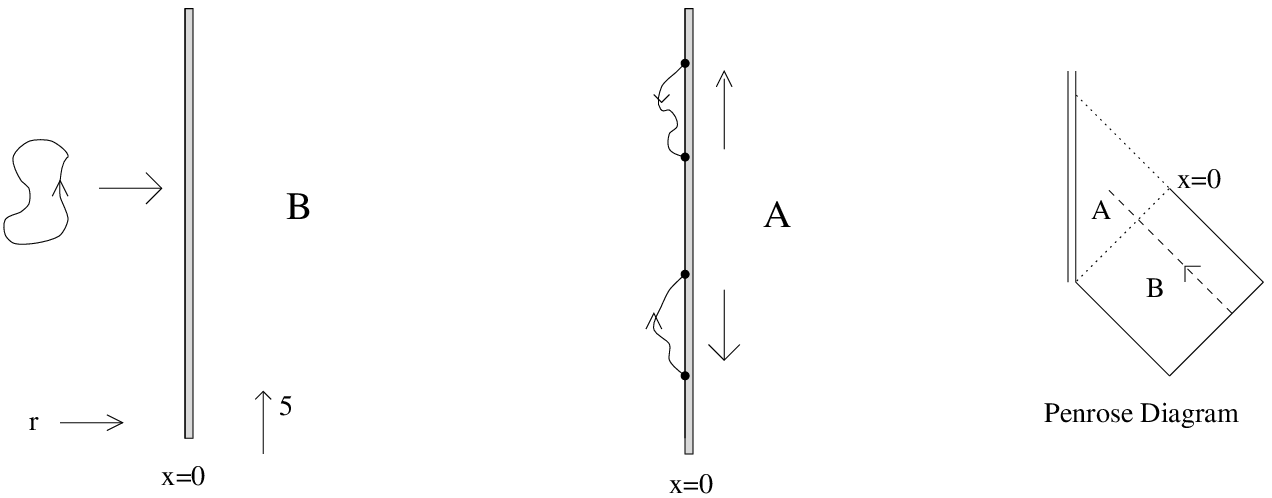}}}
{\centerline{\tenrm 
FIGURE 2: D-brane description of a  string before and after 
falling  through the horizon.
}}
}
\vskip .5cm

\newsec{Non-Extremal Black D-Branes and Hawking Radiation.}

We now turn to a discussion of nearly  extremal five-dimensional
black holes. We will find it convenient to restrict our attention to the
special black holes in which all the scalar moduli fields are constant
and the compactification geometry is completely passive. We saw that this
could be achieved in the extremal case by imposing a certain relation 
\balance\ between the three types of charge. Perturbation theory assures
us that we can achieve the same result (cancellation of the sources for
all the compactification moduli) for slightly non-extremal solutions 
by imposing the same ratio on the total source stress-energies due to 
the three different charges (we will shortly explain this is more detail).  
Since the coupling constant in these solutions is uniform in space 
we can choose it to be weak {\it everywhere}. This should be a favorable case 
for examining the non-BPS states of the D-brane, doing perturbative 
computations of their
interactions and comparing to the canonical expectations for the non-extremal 
\RN\ black hole. We will see that agreement between the two approaches
is just as impressive as in the extremal case. There is, however, a hitch: 
the presence of a large number of D-branes ($Q_1, Q_5 \gg 1$) amplifies the 
effective open string coupling constant and, in principle, renders any 
perturbative analysis of horizon dynamics unreliable
\refs{\sv,\hstro,\polchinskinotes}. We think the situation may not be so
desperate and will present some (non-rigorous!) arguments that 
open string loop corrections might not, in fact, change the essential physics. 

We perturb away from the BPS limit in a macroscopically small but 
microscopically large fashion ($ M \gg \delta M \gg $ mass of typical
excitations).  There are many ways to do this by adding stringy
excitations to the basic D-brane configuration. We are interested in those
excitations which cause the entropy to increase most rapidly with added energy.
One could add fundamental string modes traveling on the torus, but they have 
too small a central charge to be relevant. Massive open or closed string 
excitations also give a subleading contribution since the entropy of a gas 
of these excitations increases at most as $\delta M^{2/3}$, and we will
find a leading  contribution going as $\delta M^{1/2}$.
One could have five brane excitations in any direction, but that entropy
increases as $\delta M ^{p \over p+1 } $ for a membrane
of $p$ dimensions. So  we conclude  that the most
important will be the ones along the string.
There are open string  modes of type (1,1) and (5,5) starting
and ending on branes of the same type, but these modes are massive
when there is a condensate of open strings of type (1,5).
The only ones left are the modes of type (1,5) and (5,1).
If we perturb away from a purely left-moving extremal background by adding 
$\delta N_R$ right-moving oscillations, we also have to add 
$\delta N_L=\delta N_R $ left-moving oscillations to keep the total
$N= N_L-N_R$ charge fixed. These oscillations have 4 bosons and 4 fermions, 
so the central charge is the  same as it was before. The change in 
left-moving entropy is proportional to $\sqrt{ N + \delta N_R }- \sqrt N $ 
and is of order $\delta N_R/N_L$. The change in right-moving entropy is, however,
of order $\sqrt{\delta N_R/N_L}$ and dominates. 
More specifically, we find that 
$$
\left. { \Delta S \over S_e}\right|_{oscill} 
= \sqrt{ \delta N_R \over N} ~.
$$

One would like to reexpress this entropy change in terms of the 
change in mass of the configuration and compare it with the $\Delta S$
versus $\Delta M$ relation that would be extracted from the geometry
of an appropriate \RN\ solution. The question is, what solution? In our
thinking about this problem we have chosen to focus on the strict 
five-dimensional \RN\ solutions, i.e. solutions for
which the dilaton and the various compactification moduli are strictly
constant. Indeed, the extremal solution, from which we start, is of this
character because it satisfies \balance. The role of that condition is 
just to arrange that the 1-brane, 5-brane and open string
 stresses in the compactified 
dimensions add up to something which is, for instance, isotropic in the 
internal directions. If, as above, we increase the total energy carried by 
oscillations moving in the $\hat 5$ direction and change nothing else, 
the $\hat 5$ direction is singled out, compactification moduli will be 
excited and the compactification geometry will vary from point to point. 

There is nothing wrong with this, but we would like to stay within the 
framework of the strict five-dimensional \RN\ solution \reno .
Using the qualitative arguments given at the end of Sec.~2, it seems 
that to do this one should increase the source stresses due to the
1-branes and 5-branes in the same proportion as one has increased
the energy of the open strings  by introducing right-movers.
Thus we should augment the original ``branes'' wound around the internal 
directions by an appropriate number of ``anti-branes'' (for both 1-branes 
and 5-branes).  Here the word ``anti'' means having the opposite RR charge 
and opposite orientation. The anti-branes can eventually find the branes and 
annihilate, but left-movers can also annihilate right-movers, and it seems
to us that denumeration of states is equally meaningful in the two cases.
We need to add equal amounts of branes and anti-branes to keep the total 
Ramond-Ramond charges fixed. The precise quantity one should add is 
determined, in perturbation theory, by the same reasoning that determined 
the relation between the charges \balance. The logic behind that relation 
was that the condition that the dilaton not be excited 
fixes the relation of the anti-5-branes to the anti-1-branes, while
the condition that the size of the direction $\hat 5$ does not 
change fixes the relation between the branes and the oscillators.
Calling $\delta Q_1^A, \delta Q_5^A$ the extra amount of anti 
1-and-5-branes that must be added to balance an increment $\delta N_R$,
the relation which must lead to a strict five-dimensional \RN\ solution is 
\eqn\equal{
c_p \delta N_R = c_1 \delta Q_1^A = c_5 \delta Q_5^A.
}
We are assuming that the total added mass $\delta M$ is much
bigger than the mass of a single D-brane.

Just as there is an entropy increment associated with the various ways
to achieve $\delta N_R$, there are entropy increments associated with
$\delta Q_1^A$ and $\delta Q_5^A$. They are presumably independent
and should be added to get the total entropy increment. We have already 
calculated the increase due to the right and left movers, but it is not
obvious what the entropy increase due to the anti-branes will be. There is
however a U-duality \refs{\hull,\sentduality,\vafacount,\vafaintersecting}\ 
that, for example, turns anti-1-branes into right moving momentum states 
at the price of some transformation of coupling and compactification radii.
Since the entropy increase is independent of the coupling constant and the 
compactification radii, we will take the duality argument as telling us that 
the counting of the brane-antibrane states is just the same as the counting
of the left- and right-moving oscillator states. The net result for the
entropy increment is
$$
\left. { \Delta S \over S} \right|_{anti-1-branes} 
=  \sqrt{ \delta Q_1^A \over Q_1 }.
$$
Since the same argument applies to $\delta Q^A_5$,
the total entropy of the non-extremal solution is
\eqn\entropynon{  \left. {\Delta S \over S}\right|_{total} 
= \sqrt{ \delta N_R \over N} +
\sqrt{ \delta Q_1^A \over Q_1 } + \sqrt{ \delta Q_5^A \over Q_5 }
= 3 \sqrt{ \delta N_R \over N}
}
since all ratios are equal by \equal \balance . 
It is more to the point to express
this result in terms of the mass. For the extremal black hole, the mass
is just the sum of the charges (with appropriate coefficients). When we
go slightly away from extremality by adding ``anticharges'', a simple
superposition principle (valid to lowest order in $g$)  gives 
$$ M = { 1\over R }  \left\{ {c_1 \over c_p} (  Q_1 + 2 \delta Q_1^A)
+{c_5\over c_p} ( Q_5 +  2 \delta Q_5^A )
+( N + 2 \delta N_R)  \right\}~.
$$
where $\delta N_R$ represents the contribution of the added right-movers
to the mass and $N+\delta N_R$ represents the contribution of the new
total number of left-movers and so on for the other two types of charge.
Because of the ``balance'' conditions on $N$, $\delta N_R$ \balance ,
\equal , we have $\delta N_R/ N =  \delta M/2 M$ and, finally, 
$$ 
\left. { \delta S_e \over S}\right|_{string} = { 3 \over \sqrt{ 2 } }
\sqrt{\delta M \over M_e}
$$
This is the standard Bekenstein-Hawking result for the strict five-dimensional
\RN\ black hole, with the correct normalization and functional dependence 
on mass. Although the arguments that led us here are less than rigorous
(especially the duality argument for entropies associated with branes and 
antibranes), the simple end result gives us some confidence in the intermediate
steps. If we had studied the deviation from extremality arising from adding
only the right-moving momentum states (thus moving away from a strict
five-dimensional solution), the entropy increment formula would simply have
been missing the factor of three, presumably the correct result for that
state \hstro.

These non-BPS states will decay. The simplest decay process is
a collision of a right moving string excitation with a
left moving one to give a closed string that leaves the
brane. We will calculate the  emission rate for 
 chargeless particles, so that the basic process is a 
 right moving open string with momentum $p_5 = n/R$ colliding 
with a left moving one of momentum $p_5 = - n/R$ to give a closed
string of energy $k_0 = 2 n/R $.
We will calculate the rate for this process according to
the usual rules of relativistic quantum mechanics and show
that the radiation has a thermal spectrum if we do not
know the initial microscopic state of the black hole.

The state of the black hole  is specified by giving the
left and right moving occupation numbers of each of the
$ 4 Q_1 Q_5$ bosonic and fermionic oscillators.
In fact, the nearly extremal black holes live in
a subsector of the total Hilbert space that is
isomorphic to  the Hilbert space of a one dimensional
gas of massless particles of $ 4 Q_1 Q_5$ different types,
either  bosons or  fermions (up to subtleties related
to the orbifolding procedure  discussed above).
This state $|\Psi_i\rangle$ can then emit  a
closed string and become $|\Psi_f\rangle$.
The rate, averaged over initial states and summed over final states,
as one would do for calculating the decay rate of an unpolarized
atom, is
$$
d \Gamma \sim  { d^4  k \over k_0 } { 1 \over p_0^R p_0^L }
\delta( k_0 -( p_0^R + p_0^L) ) \sum_{i,f}
\left| \langle \Psi_f| H_{int} | \Psi_i \rangle  \right|^2
$$

The relevant string amplitude for this process is given
by a correlation function on the disc with two insertions
on the boundary, corresponding to the two open string
states and an insertion in the interior, corresponding to
the closed string state (in the spirit of \scattering ).
 The boundary vertex operators
change boundary conditions for four of the coordinates.
We consider the case when the outgoing closed string is
a spin zero  boson in five dimensions,
so that it corresponds to the dilaton,  the
internal metric,  internal $B_{\mu\nu}$ fields,
or internal components of RR gauge fields.
This disc  amplitude, call it  $\cal A $,  is proportional to the
string coupling constant $g$  
and it is basically 
 the same amplitude that
that would appear if we were to calculate the
absorption cross section of the black hole.

Note that performing the average over initial and sum over final states
will just produce a factor of the form $ \rho_L(n) \rho_R (n)$ with
\eqn\rhoright{
\rho_R (n) = {1 \over N_i} \sum_{i} \langle \Psi_i | a^{R \dagger}_n
 a^R_n | \Psi_i
\rangle
}
where $N_i$ is the total number of initial states and
$ a_n^R $ is the creation operator for one of the $ 4 Q_1 Q_4 $
bosonic open
string states. The factor $ \rho_L(n) $ is similar.
Since we are  just
averaging
 over all possible initial states with given value of $\delta N_R$,
this corresponds to taking the expectation value of $a^\dagger_n a_n$ in
the microcanonical ensemble with total energy $\delta N_R$ of a one
dimensional gas. Because  $\delta N_R$ is large compared to one (but still
much smaller than $N_L$), we can calculate \rhoright\ in 
the canonical ensemble. Writing the partition function as
$$ 
Z = \sum_N q^N d(N) = \sum_N  q^N e^{ 2 \pi \sqrt{ Q_1  Q_5 N} },
$$
doing a saddle point evaluation and then determining $q$ from
$$
\delta N_R = q {\p \over \p q } \log Z,
$$
we find $ \log q = - \pi \sqrt{Q_1 Q_5 / \delta N_R}  $.
Then we can calculate the occupation number of that level as
$$ \rho_R(k_0) = { q^n \over 1-q^n } = { e^{- k_0 \over T } \over
1 - e^{ - k_0 \over T } }
$$
We can read off the ``right moving'' temperature
$$ T_R = {2 \over \pi}   {1 \over R }
 \sqrt{ \delta N_R \over Q_1 Q_5 } 
$$
Now using \entropyd ,~  \coeff ~and \balance ~
 we get 
\eqn\hawkingstring{
T_R = { 2 \over \pi r_e } \sqrt { \delta M \over 2 M }
}
There is a similar factor for the left movers $\rho_L$
with a similar looking temperature
$$ T_L = {2  \over \pi}   {1 \over R }
 \sqrt{  N \over Q_1 Q_5 }
$$
Since  $T_R \ll T_L$ the typical energy of the
outgoing string will be $k_0 \sim T_R $ and
$k_0/ T_L \sim T_R/T_L \ll 1 $ so that
we could approximate
\eqn\rholeft{
 \rho_L \sim { T_L \over k_0  }=
{ 2  \over \pi k_0 R }
\sqrt{ N \over Q_1 Q_5}
}
The expression for the rate then is, up to a numerical constant
\eqn\ratenew{
d\Gamma   \sim  { d^4 k \over k_0 } { 1 \over p_0^R p_0^L }
|{ \cal A}|^2  Q_1 Q_5 \rho_R(k_0)  \rho_L(k_0)
}
where ${\cal A}$ is the disc diagram result.
 The factor
 $Q_1 Q_5$ is due to the fact that we have this many distinct
open string levels for each momentum $n$.
The final expression for the rate is then, using 
\rholeft ~in \ratenew 
\eqn\ratefinal{ d\Gamma   \sim
({\rm Area} )  { e^{- k_0 \over T_R } \over
1 - e^{ -k_0 \over T_R } } d^4 k 
}
We conclude that the emission is thermal, with a physical 
Hawking temperature
\eqn\hawkingstring{
 T_H = T_R =  { 2 \over \pi r_e } \sqrt { \delta M \over 2 M }
}
which exactly matches the classical result \hawking .
It is an interesting  result that the area appeared correctly in
\ratefinal . Actually,   the coupling constant coming from
the string amplitude $\cal A$   is hidden in the
expression for the area. 
Of course, it will be very interesting
to calculate the   coefficient  in \ratefinal ~to
see whether it exactly matches the absorption coefficient
 of a large 
classical black hole. In fact the result should not
depend on the internal polarizations of the outgoing particles.
Note that   U-duality \hull  ~suggests  that
the  anti-brane excitations will also produce Hawking radiation
with the same temperature \hawkingstring  .
Notice that if we were emitting a spacetime fermion then 
the left moving string could be a boson and the right moving
string a fermion, this produces the correct thermal factor
for a spacetime fermion. The opposite possibility gives
a much lower rate, since we do not have the enhancement
due to the large $\rho_L$ \rholeft . Notice also that
when separation from extremality is very small, then the number
of right movers is small and the statistical arguments used
to derive \ratefinal\ fail. Such deviations  were  observed in 
\limitations . 

We now examine the range of validity of these approximations.
For the purposes of this argument, we take the compactification
radii to be of order $\alpha'$ and set  $\alpha'=1$. In this
case \balance ~implies $Q_1\sim Q_5\equiv Q$ and $Q \sim g N$. 
Then, by \entropyext, we find that the area of the horizon is 
$A\sim(g^2 N)^{3/2}$. In order for the classical black hole interpretation to
hold, this area has to be much larger than $\alpha'$, so $g^2 N \gg 1$.
Since we want  $g$ to be  small,  $N$ is very large. This seems to 
invalidate the perturbative D-brane picture since
open string loop corrections are of order $g Q = g^2 N $, and, 
due to the large number $Q$  of D-branes, they are likely to be large 
\refs{\sv , \hstro , \polchinskinotes }. We will try to argue that 
open string corrections might in fact be suppressed. Let us first 
give a very qualitative argument. 
It has been noted \hologram\ that string perturbation
theory is expected to  break down 
when the number of strings per unit volume
is of order $1/g^2$. In this case we have of the order
$N$ open strings sitting at $x=0$, but the area of this ``point''
has been expanded so that now the density of strings is
$ N/A_H = N/(g^2 N)^{3/2} \ll 1/g^2 $ in the classical black hole 
domain ($ g^2 N \gg 1 $).
Going back to the open string loops, we note that the loop will
be in a  non-trivial  background of open strings. In fact, this 
background was crucial to obtain a small
five dimensional coupling constant and
 non-zero area, which implies that somehow the D-branes might be  
``separated'' from each other. 
We suspect that this background of open strings suppresses
open string loops, enabling us to get results off extremality. 
This is of course something to be checked in detail. 
It is clear, however, that there are some circumstances where
open string backgrounds suppress loop contributions. For
example, compare loop contributions of $n$ D-branes on top of
each other  and $n$ widely separated D-branes. The difference
is just a background of open string translational zero modes. 

Finally, the fact that the perturbative D-brane treatment of non-extremal
physics gives the right results strongly suggests that there is more
than a grain of truth here. We think it quite possible that open string
quantum corrections are not as large as suggested by naive estimates.
The skeptic is  entitled to disagree! 

\newsec{Summary, Conclusions and Speculations.}
Our goal was to find a black hole which could simultaneously be
described as a solution of low-energy effective field theory (so that
the usual results of general relativity would be applicable) and as a
weakly coupled D-brane soliton (so that perturbative string theory
could be used to count states and evaluate entropies). Such an
object would make possible a serious confrontation between string theory
and the paradoxes of black hole quantum mechanics. We come close to what we
need in certain solutions to ten-dimensional supergravity, compactified to
five dimensions by wrapping 5D-branes around a $T^5$ combined with 1D-branes
winding in one of the $T^5$ directions and and a gas of open strings carrying
momentum.  This object has three charges: the charge of the 5D-branes,
the charge of the 1D-brane and the longitudinal momentum along the
direction of the 1D-brane. When all three are excited,
there is a horizon and nothing singular happens either to the dilaton
or to the volume of the $T^5$ at the horizon: from the point of view
of low-energy physics everywhere outside the horizon, this is
a large, classical  five-dimensional \RN\  black hole. For
appropriate choices of parameters, it can be made extremal and BPS-saturated
(i.e. it preserves some supersymmetries). This solution has
an event horizon with finite area, and therefore is expected to have an
entropy, but it is hard to give a classical description of
the states responsible for the entropy.

In the D-brane picture these states are
easily denumerated and they lead to an entropy
which agrees perfectly with that calculated from the horizon area of the
corresponding supergravity solution. It is easy to construct a non-BPS
saturated state by mixing in some right-moving open strings with the
left-movers: these states are also denumerable and the increase
of the entropy as the mass is increased above the extremal value can
be computed. There are other states, arising from mixing in one-branes
and five-branes with opposite winding from the background branes, but
U-duality \hull\ suggests  how to count these states (how
to do this state counting would be a total mystery without D-branes). The
non-extremal entropy can be accounted for and the decay rate shows a thermal
distribution with the ``Hawking'' temperature, in perfect agreement with the
canonical formulae \hawkingentropy. The precise mechanism by which
non-extremal D-branes decay is apparent: two open strings of opposite
longitudinal momentum annihilate to create a massless closed string
state which escapes. The resulting radiation rate is
proportional to the horizon area, but it has yet to be demonstrated
that the overall numerical coefficient is that of a black body for all
channels. For non-extremal black holes, these pleasing results are
undercut by the fact that the large number of D-branes seem to render
 the horizon
open strings strongly coupled. We gave qualitative arguments that the
open string loops should in fact be suppressed, but they need to be 
developed further to be convincing.

What does all this mean? 
Assuming that loop corrections are under control,
it means that unitary evolution 
of quantum states is {\it not} violated by this type of black hole. 
Entropy would be  just the degeneracy of an identifiable set of states that
live on the horizon. Thermal emission would be  due to the lack of
knowledge of the initial state, just as in real-world thermodynamics.
There would be  no information loss: quantum states falling into the black
hole from the outside would cause a unitary evolution in the Hilbert space
 of horizon states that ``records'' the infalling quantum
information. The remarkable novelty of the D-brane approach is that it 
tells us exactly how to calculate amplitudes for transition between
open-string horizon states as external closed string states are emitted,
absorbed or scattered: they are the familiar disc amplitudes, responsible
for coupling between open and closed strings, which were much studied
in type-I string theory. This direct access to the dynamics
of horizon states is what has been the missing ingredient in discussions
of the black hole information loss problem. 

In short, we have in our hands an extremely simple model black hole in
which physically very nontrivial calculations of horizon quantum dynamics 
seem to reduce to concrete string theory calculations. This should
permit us to see how and whether the speculative mechanisms for resolving
black hole paradoxes get realized. Three topics one might mention are
the membrane  paradigm \membrane, complementarity \complementarity\ and 
the holographic principle \hologram.
It is clear that the open strings provide a description of the state of a 
membrane on the horizon. It is clear that, from the D-brane perspective,
there is nothing for an infalling closed string to do, apart from
reflecting back to the outside world, except become an open string on
the horizon. These horizon {\it surface} degrees of freedom must encode the
state of ordinary strings that, in the usual spacetime picture, have
fallen into the world on the other side of the horizon. To implement
complementarity, there must be some  non-local transformation 
between the horizon open strings and effective closed strings describing
the measurements accessible to an infalling observer who has fallen inside
the horizon. To explore all this, and to better understand how string theory 
resolves the black hole paradoxes, one urgent task is to provide a good 
estimate of loop corrections in this background of open strings. 
It would also be interesting to try to find the mapping 
between the usual spacetime description of geodesics crossing the horizon
and the three-plet of discrete numbers $(i,j,n)$ ($i,j$ label the
particular branes) that is used to index the open string  D-brane states. 
It would be very nice to find such a simple, and explicitly calculable,
black hole system in four dimensions, though the essential qualitative 
features are probably already displayed in the five dimensional
solution.

\newsec{Acknowledgements}

We thank specially I. Klebanov and F. Larsen for valuable help 
and comments. We also benefited from discussions with A. Hashimoto,
A. Peet and C. Schmidhuber. The original bulletin board submission
of this paper suffered from a serious underestimate on our part of the 
severity of the open string strong coupling problem (coupled with
a corresponding overestimate of the strength of our conclusions concerning 
the notorious problems of black hole quantum mechanics). We want to thank
several correspondents, most notably A. Strominger, L. Susskind and
C. Vafa for their prompt and helpful criticisms which we have used to,
we hope, improve this modified submission. 

This work was supported in part by DOE grant DE-FG02-91ER40671.


\listrefs

\bye